# Half-metallicity and anomalous Slater-Pauling behaviour in half-Heusler CrMnSb

Himanshu Joshi[1*], Shradhanjali Dewan[1], Lalrin Kima[2,3], Aldrin Lalremtluanga[2], Homnath Luitel[4], K. C. Bhamu[5] and D.P. Rai[2*]

[1] *Department of Physics, SRM University Sikkim, Gangtok, 737102 India*

[2] *Department of Physics, Mizoram University, Aizawl-796004, India*

[3] *Physical Sciences Research Centre (PSRC), Department of Physics, Pachhunga University College, Aizawl-796001, India*

[4] *Department of Physics, Nar Bahadur Bhandari Government College, Sikkim, Gangtok, 737102 India*

[5] *Department of Physics, Mody University of Science and Technology, Laxmangarh, Rajasthan, 332311 India*

*Corresponding authors: Himanshu Joshi (himanshuabijoshi09@gmail.com), D. P. Rai (dibyaprakashrai@gmail.com),

**Abstract:** This study provides a first-principles insight into half-Heusler CrMnSb to understand its deviation from the conventional Slater-Pauling semiconducting behaviour. CrMnSb having a valence electron count (VEC) of 18, has been proposed to exhibit compensated ferrimagnetic character instead of the expected nonmagnetic semiconducting ground state. As half-Heusler system with VEC = 18 is not known to exhibit magnetic ordering, we have investigated the details of the electronic and magnetic properties of CrMnSb using a combination of density functional theory and Green's function-based multiple-scattering theory. We show that, despite satisfying the 18-valence electron Slater–Pauling rule, CrMnSb does not exhibit ground-state nonmagnetic semiconducting behaviour. Instead, it reveals a half-metallic, fully compensated ferrimagnetic ground state. This anomaly is found to originate from the presence of localized sublattice moments, resulting from antiparallel alignment between Cr and Mn sublattices, enforces half-metallic ferrimagnetism despite its ideal 18 valence electron count.

# 1. INTRODUCTION

Heusler compounds were first discovered in 1903 by Friedrich Heusler, who reported ferromagnetic behaviour in Cu$_2$MnAl, despite the absence of inherently magnetic elements in its composition [1-3]. This unexpected property led extensive research in these types of material, leading to the identification of over 1500 Heusler intermetallic systems. These compounds demonstrate a wide range of functional properties, and are found to have diverse applications in the field of spintronics, magneto-electronics and thermoelectric devices [4-7], due to their ability to exhibit semiconducting, ferroelectric, superconducting and topologically non-trivial behaviour [8-10]. Structurally, Heusler compounds are generally classified into two principal types based on their stoichiometry. The full-Heusler (FH) alloys follow the formula X$_2$YZ, while half-Heusler (HH) are represented by XYZ, with X and Y typically being transition metals, while Z a main-group element [11]. Their electronic and magnetic characteristics can often be anticipated using the Slater-Pauling rule, which relates the total magnetic moment to the number of valence electrons ($Z_t$) [11]. For HH alloys, the rule is given as $M = Z_t - 18$, and for FH alloys, as $M = Z_t - 24$, where M is the magnetic moment per formula unit [12]. Consequently, HH with valence electron count (VEC) of 18 are typically expected to be non-magnetic semiconductors according to this rule [13]. Thus, CrMnSb, with VEC of 18 should ideally be a non-magnetic semiconductor, however we report the material to exhibit half-metallic characteristics, with net zero magnetic moment.

The density of states (DOS) computed using both GGA and GGA+U methods show metallic characteristic in majority spin channel, while minority spin channel exhibits semiconducting nature, revealing the half-metallic property of the compound. Recent studies on 18 valence HH [14-17] have also been found to show similar characteristics. These materials are reported to exhibit ferrimagnetic behaviour, which is rarely observed in 18 VEC HH compounds, whereas conventionally they should have equal DOS contribution to the spin up and spin down channels, making them paramagnetic. HH other than 18 VEC, usually have asymmetric contribution to the spin up and spin down channels making them ferromagnetic in nature, whereas some reports also predict antiferromagnetic ordering [18-22]. Recent theoretical and computational studies [23-25] further suggest that ferrimagnetic structure could stabilize in 18 (for HH) and 24 (for FH) valence electron Heusler, though experimental validation remains limited.

We report CrMnSb to show fully compensated half-metallic behaviour characterized by complete spin polarization and no net magnetic moment. The absence of net magnetisation is suggested to be particularly advantageous for spintronic applications when compared to other half-metals, as it minimizes stray magnetic fields and supresses the formation of magnetic domains [26-28]. Thus, such materials are in particular demand for memory devices. However, the fully compensated characteristic in CrMnSb is still under scrutiny, with studies suggesting 50% substitution of Sb by P is necessary to fine tune the electronic and magnetic balance [17]. Moreover, the compound is known to crystallize in more than one structural phase, leading to conflicting reports [29]. Nevertheless, our investigation focuses on explaining the half-metallic behaviour of system, predicted to be semiconducting. We note that the magnetic compensation and electronic structure are extremely sensitive to the lattice parameter, where even a slight expansion disrupts the net-zero magnetization. We show through calculated exchange coupling parameters that Cr and Mn magnetic moments govern the magnetic behaviour in CrMnSb. Additionally, X-ray Magnetic Circular Dichroism (XMCD) plots confirm the opposite alignment of Mn and Cr magnetic moments, which ultimately leads to ferrimagnetic behaviour in the material.

## 2. COMPUTATIONAL DETAILS

The full-potential linearized augmented plane wave method (FP-LAPW) [30] based Wien2k code [31], employing density functional theory (DFT) [32, 33], forms the basis of calculation in this work. Considering the discrepancies reported [34, 35] in generalized gradient approximation (GGA) [36] to effectively capture the electronic structure due to unrealistic self-coulomb repulsion, the GGA+U method [37] is additionally utilized as the correction over GGA. The effective Coulomb potential, $U_{eff} = U_{Cr/Mn} - J_{Cr/Mn}$ with interaction parameters $U_{Cr} = 1.5$ eV, $U_{Mn} = 3.5$ eV and $J_{Cr/Mn} = 0$ was considered. In0side the muffin-tin spheres, the charge density and potential include contributions from non-spherical components, expanded up to a maximum angular momentum quantum number of $l_{max} = 10$. The cutoff condition for the plane wave basis was defined by the product $R_{MT} \times K_{max} = 7$, where $R_{MT}$ is the smallest muffin-tin radius in the unit cell and $K_{max}$ the maximum reciprocal lattice vector. In the interstitial regions, both the charge density and potential are expanded using a Fourier series with $G_{max} = 12$. To achieve reliable self-consistent field convergence, the energy tolerance was restricted to $10^{-5}$ Ry, and a dense $20 \times 20 \times 20$ k-point mesh was employed to sample the Brillouin zone. The exchange coupling constants between the atomic pairs were computed using the Green's function-based multiple-scattering formalism [38] as implemented in the Spin-Polarized Relativistic Korringa Kohn Rostoker (SPR-KKR) code [39]. This method enables a parameter-free evaluation of exchange coupling constants.

## 3. RESULTS AND DISCUSSION

CrMnSb crystallizes in a cubic structure with two distinct phases, each with different vacant atomic site. The α-phase has a vacant 4b Wyckoff site, whereas the γ-phase has at 4d [29]. In this work, we focus on the α-phase with optimized lattice constant 5.87 Å and F-43m (No. 216) space group. Within the unit cell, the Cr atoms are at the corner of the cube, while Mn and Sb form a parallel plane to Cr layer, leaving the central site of the cube unoccupied. In this arrangement, the Cr and Mn moment are aligned anti-parallel to each other (Figure 1). Previous studies suggest the γ-phase of this compound to be fully half-metallic, when structural optimization were optimized without considering spin polarized calculations [17]. This causes the unit cell to undergo isotropic compression into a metastable state with lattice constant ~ 5.8 Å. Structural optimization with spin polarized calculation leads to higher values of lattice constant ~5.98 Å and thus fail to produce a net zero magnetic moment. This inconsistency has cast doubt on the fully compensated magnetism in the γ-phase. To avoid this ambiguity, we adopt the α-phase optimized with spin polarization to clearly understand the compounds intrinsic magnetic behaviour.

The electronic structure obtained within GGA approach is shown in figure 2 and 3. The total and elemental resolved density of states confirms the half metallic nature with energy gap in the minority spin. CrMnSb displays complete magnetic compensation and possesses a spin-selective energy gap of 0.93 eV in the minority spin channel, which is slightly higher than the available literature [17, 29]. The difference in the band gap value can likely be attributed to variations in the lattice parameters employed during the calculations, as well as the methodological differences between the FP-LAPW approach used here and the pseudopotential-based method adopted in reported studies. From the DOS plots, in the spin-up channel, Cr and Mn atoms have significant contribution at the Fermi level, leading to a metallic state. In contrast, DOS at the Fermi level vanishes in the spin-down channel, indicating spin selective semiconducting characteristic due to the observed energy gap. Thus, full spin polarization occurs at the Fermi, which is a signature of half-metallicity.

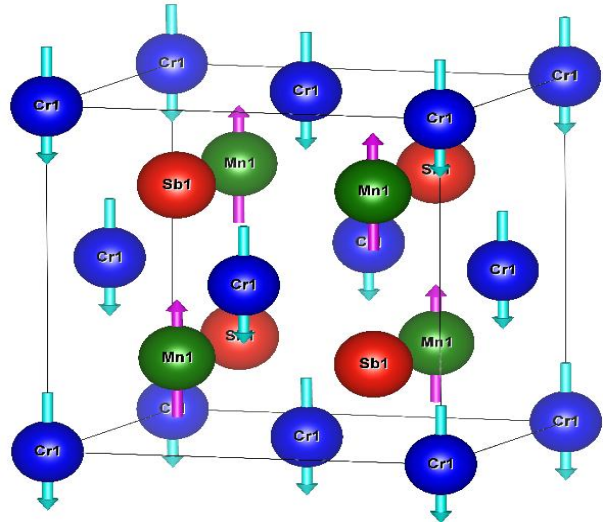

**Figure 1**: Crystal structure of CrMnSb with anti-parallel coupling between Mn-Cr atom.

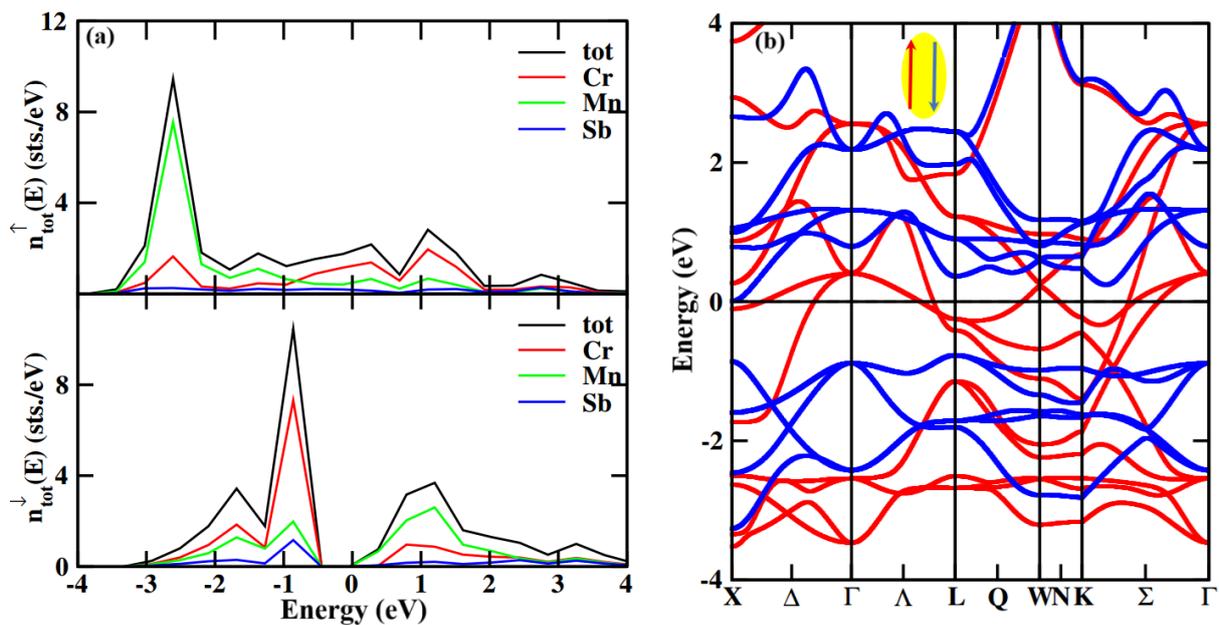

**Figure 2**: **(a)** Density of States (DOS) with elemental contribution and **(b)** band structure of CrMnSb from GGA.

The spin resolved partial DOS (PDOS) reveal the lower energy region near -4 eV has contributions from Sb -$p$ states in the both spin channels (Figure 3a). However, deeper valence states near -10 eV (energy range not included in the figure), contributions arise from Sb -$s$ state. The sharp DOS peaks in the spin-up valence region between -2 eV to -3 eV arises from strong Mn -$d$ states, while similar sharp peak in the spin down channel is due to Cr -$d$ states. The orbital resolved DOS (Figure 3 b) shows the Cr and Mn -$d$ states split into $t_{2g}$ and $e_g$ components due to crystal field effect. The Mn $t_{2g}$ states dominate the spin-up channel, whereas Cr $e_g$ states are more prominent in the spin down region. Since the positive spin channel is dominated by

Mn contribution, hence it has a positive magnetic moment and as spin-down channel is dominated by Cr, its magnetic moment is negative. This asymmetry confirms different sublattice moments, a fundamental characteristic of ferrimagnetism. Additionally, a minor Sb – $p$ state contribution in the spin down channel introduces to a small negative magnetic moment on the Sb site. The individual atomic magnetic moments tabulated in Table 1, reveal that the Cr and Mn moments nearly cancel each other, with minimal contributions from the interstitial and Sb atoms. As a result, the total magnetic moment of the unit cell approaches zero, confirming a fully compensated magnetic state. The combination of spin and orbital resolved DOS features along with net-zero magnetisation identifies CrMnSb as a fully compensated ferrimagnet. The magnetic moment contribution from the interstitial reason is due to the FP-LAPW approach of calculation and accounts for the electron spin density in the space between atomic spheres, arising from delocalized electrons. Their values are typically small but contributes to the total magnetic moment.

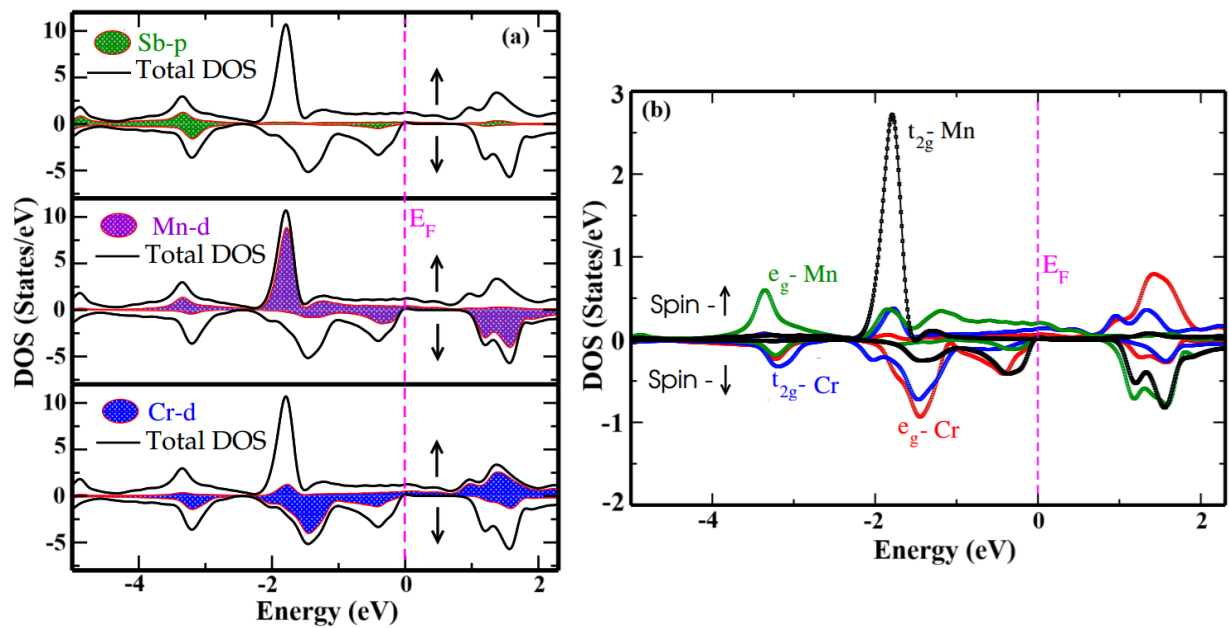

**Figure 3**: **(a)** Spin resolved Partial Density of States (PDOS) and **(b)** Orbital resolved DOS from GGA

**Table 1.** Calculated magnetic moments, exchange splitting and energy band gaps. IR in the table represents interstitial region.

| Functionals | Magnetic Moments (μB) | | | | | Exchange splitting Δ (eV) | | Band Gap (eV) |
|---|---|---|---|---|---|---|---|---|
| **Atoms** | Cr | Mn | Sb | IR | Total | Cr | Mn | $E_g$ (↓) |
| GGA | -2.48557 | 2.71322 | -0.03148 | -0.19358 | -0.0001 | 8.65 | 10.86 | 0.93 |
| GGA+U | -2.74760 | 2.96879 | -0.03013 | -0.19112 | 0.0002 | 8.79 | 11.13 | 1.19 |

The electronic structure calculated with Hubbard U correction (Figure 4), exhibits notable modification in the DOS characteristics, whereas the overall fully compensated ferrimagnetic character and half-metallic nature is preserved. A noticeable change is in the increased magnitude of Mn and Cr magnetic moments, attributed to enhanced localization of *d*-electrons, implicating stronger anti-parallel alignment between Cr and Mn atom. The spin-down channel band gap widens by approximately 28 % reaching a value of 1.19 eV, as self-interaction error of GGA is corrected, leading to more accurate treatment of electron correlations. This results to improved spin polarization, reinforcing the materials half-metallic characteristic [40]. The exchange splitting is more pronounced and the DOS near Fermi level develops more intense features, indicating that the gap opening is driven by improved electron localization. The GGA+U calculation revels reduced band dispersion near the Γ-point, highlighting enhanced quasiparticle effective mass. This behaviour reflects the role of U correction in localizing electrons and amplifying correlation effect leading to band renormalization.

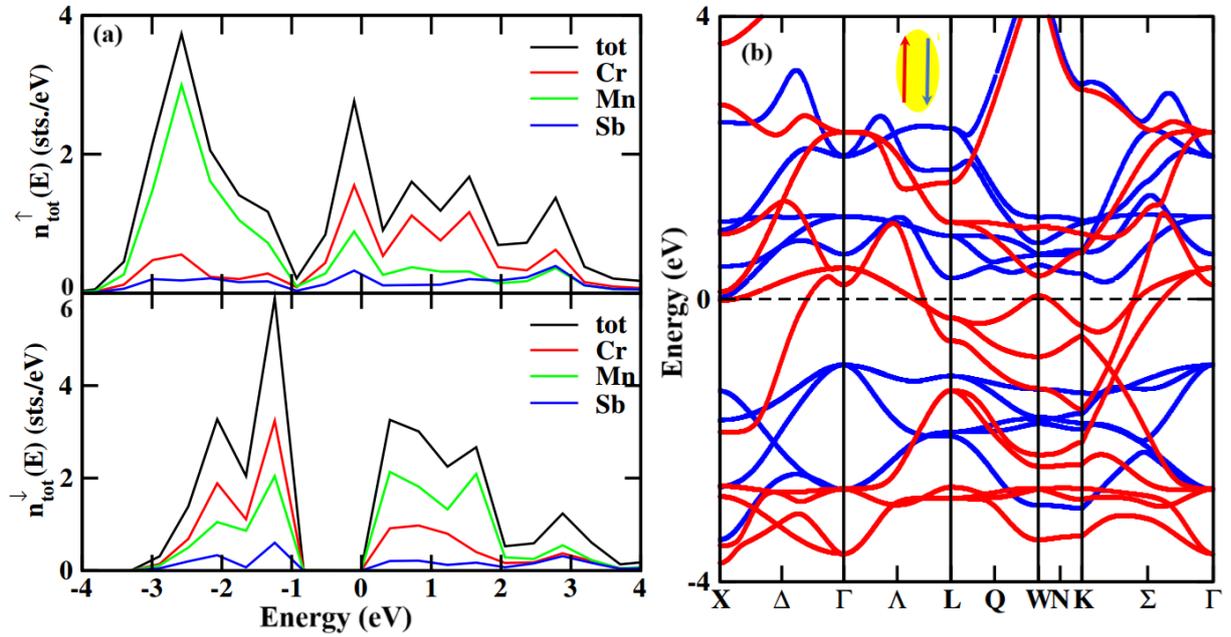

**Figure 4**: **(a)** Density of States (DOS) with elemental contribution and **(b)** band structure of CrMnSb from GGA+U.

The significance of GGA+U method in capturing electronic correlations, important for spintronic applications can be further analysed and verified from the X-ray Absorption Spectra (XAS) and X-ray Magnetic Circular Dichroism (XMCD) plots (Figure 5). The $L_3$ and $L_2$ peak shows significant shift with the inclusion of the Hubbard U correction. $L_3$ peaks correspond to the transition from the core $2p_{3/2}$ states to the 3*d* unoccupied state, whereas $L_2$ peaks correspond to the transition from $2p_{1/2}$ →3*d*. The XAS plot shows almost equal intensities for Cr and Mn atom, indicating a large number of unoccupied 3*d*-states. The notable spin-orbit coupling in the 2p core levels leads to a distinct energy difference between the $L_3$ and $L_2$ edges. Since the 3*d* states are spin-polarized, examining the XAS line shape, especially through XMCD, offers essential element-specific information regarding the magnetic characteristics of the Cr and Mn atoms in the CrMnSb compound. The opposite direction of the Cr and Mn peaks (Figure 5 a, b) in the XMCD spectra corresponds to the antiparallel spin alignment and is consistent with the DOS results. This opposite sign confirms the ferrimagnetic coupling between the Cr and

Mn spin. The GGA based XMCD spectra shows a clear single peak centred at the $L_2$ and $L_3$ edges in Mn. Notably, a clear shoulder structure appears near the $L_3$ edge with GGA+U. These results suggest the localized character of the Mn 3d states due to improved treatment of electron correlations and localization offered by U correction. Similar analysis can be applied for the shoulders appearing near $L_2$ peak in Cr. The exchange splitting energy also changes significantly due to the shift in $L_{2,3}$ edges. The origination of the shoulder peak can be attributed to the inter-band transition from $2p$ to unoccupied $3d$ states in Cr and Mn atoms [41, 42]. In contrary to the DOS analysis, the shoulder feature results from the significant DOS peaks near the fermi level, which therefore becomes more pronounced in GGA+U. The overall intensity of the observed XAS and XMCD spectra remains almost similar with both GGA and GGA+U, however a dominant negative peak is observed for Mn near the $L_3$ edge which is due to the unoccupied density near the Fermi level. This reflects transition into minority spin state due to helicity of circularly polarized photon with the spin orientation and does not imply a negative moment [43,44]. The enhanced spin-orbit effects due to corrected electron correlation are evident from the additional shoulder features arising with GGA+U.

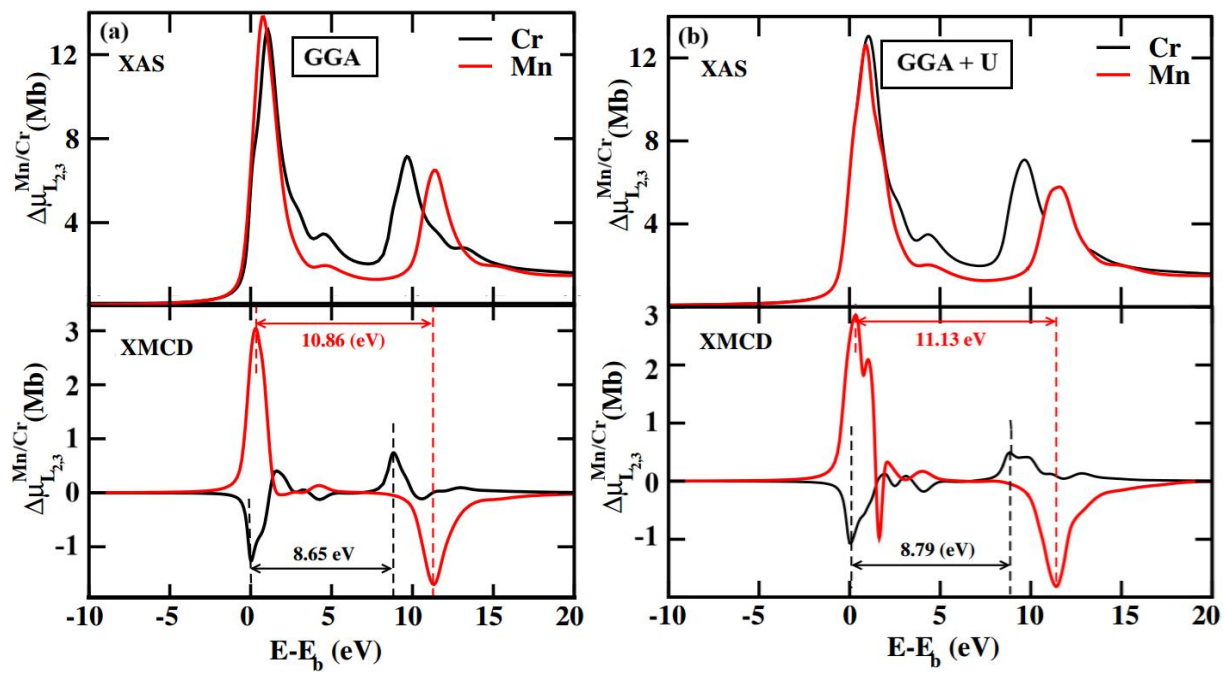

**Figure 5**: X-ray Absorption Spectroscopy (XAS) and X-ray Magnetic Circular Dichroism (XMCD) spectra at the $L_{2,3}$ edges of Cr (black) and Mn (red) in CrMnSb, calculated using (**a**) GGA and (**b**) GGA+U methods.

In figure 6, we show the atomic pairwise exchange interaction ($J_{ij}$) as a function of interatomic distance obtained from GGA+U method. This exchange parameter is essential to understand the microscopic origin of ferrimagnetic character in CrMnSb and also to understand the XMCD spectra. The figure shows the exchange interaction importantly only between the first nearest neighbours. The interactions that do not appear are negligible or very small. The dominant exchange coupling occurs from Cr and Mn interaction with a large negative $J_{ij}$ value and indicates strong antiferromagnetic (AMF) coupling between the two sublattices [45]. In contrast, the Cr–Cr and Mn–Mn intra sublattice interactions are much weaker and predominantly positive, suggesting ferromagnetic (FM) alignment within each sublattice. The Cr-Sb and Mn-Sb interactions are very weak and non-magnetic. The strong AMF coupling between Cr-Mn atom (~ -24 meV) forces their spin into anti-parallel alignment with Mn spin up (↑) and Cr spin down (↓), also evident from the XMCD plots. The Mn-Mn and Cr-Cr intra-sublattice FM interactions enforces uniform spin orientation within each sublattices, stabilizing the overall spin configuration, thus reinforcing AFM alignment across the sublattices. The decay in $J_{ij}$ after a short range further indicates minimal magnetic frustration. Altogether, the resulting spin configuration balances the magnetic moments, yielding net-zero magnetization.

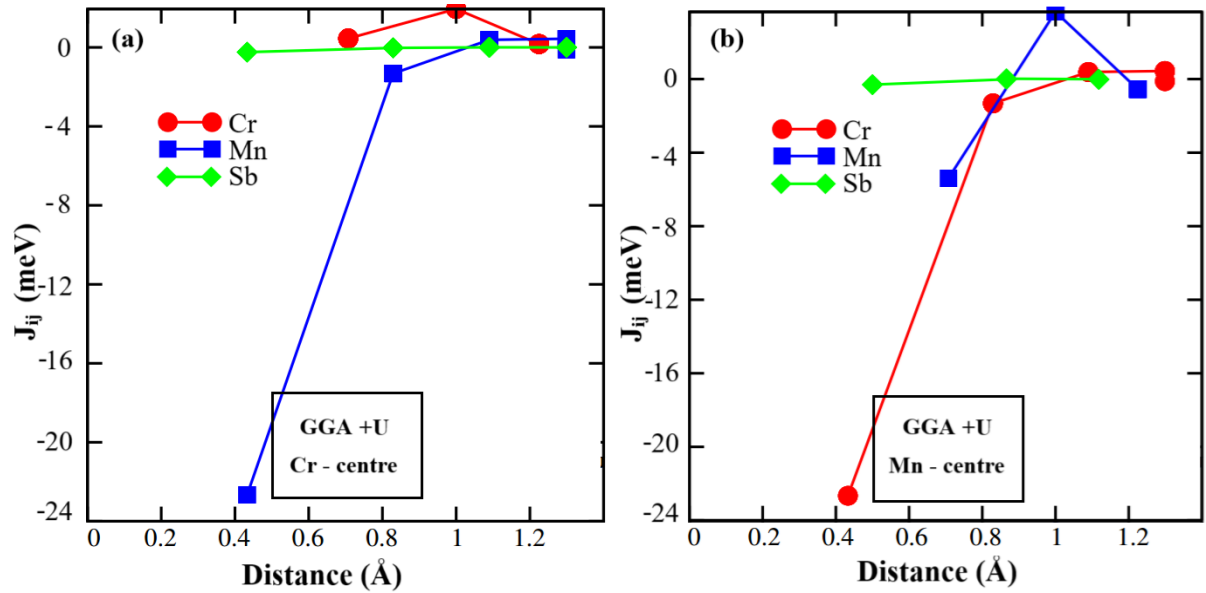

**Figure 6**: Exchange coupling parameter as a function of atomic distance with GGA +U **(a)** Cr-atom centre and **(b)** Mn-atom centre.

The DOS plots (Figure 2, 3, 4) along with the available literature [17, 29] suggest strong Sb-p hybridization with Mn-d and Cr-d states. Also, the exchange coupling plots evidently confirms Sb remains non-magnetic. This strongly indicates, the AMF coupling between Cr and Mn atom is mediated indirectly by Sb via super-exchange pathway mechanism [46-48]. The AFM coupling between Cr and Mn atom is indirectly facilitated by the Sb atom orbitals. This effectively acts as a bridge in the exchange pathway. An electron from Sb *p*-orbital channel towards Cr *d*-orbital, matching Cr (↓) spin orientation. This creates a hole in Sb with opposite spin (↑), which is filled by electron from Mn-*d* orbital, aligned with hole spin (↑). Thus, as a result, Cr acquires ↓ spin and Mn acquires ↑ spin, thereby establishing the observed antiparallel configuration between the two atoms. Hence, Sb mediates the AFM interaction in Cr and Mn atom, without contributing significant magnetic moment itself.

### 4. Conclusion

We report an in-depth analysis on the electronic and magnetic characteristic of half-Heusler CrMnSb using first principles method based on the FP-LAPW approach, within density functional theory. To accurately capture the electron correlation and exchange interaction, the Hubbard U correction was employed to the GGA functional for more reliable description of the electronic band structure and magnetic states. We considered the α-phase of the compound optimized with spin polarization calculation. CrMnSb, satisfies the Slater–Pauling criterion of 18 valence electrons per formula unit and is therefore typically predicted to be a non-magnetic semiconductor. However, the compound instead exhibits a half-metallic, fully compensated ferrimagnetic character. We explained this deviation by investigating the underlying electronic and magnetic properties of the compound.

The DOS plots in figure 2,3 and 4(a) clearly shows spin dependent electronic structure, with metallic majority spin-up channels and semiconducting minority spin-down channel. The observed band gap within GGA+U formalism in the semiconducting spin-down channel was 1.19 eV. The asymmetric spin contribution in the DOS highlights the half-metallic character in the compound. Mn $t_{2g}$ states dominate the spin-up channel, whereas Cr $e_g$ states are more prominent in the spin down region. Thus, an orbital resolved asymmetry is observed between the two transition metals and in comparison with the calculated magnetic moments in Table 1, the anti-parallel spin alignment in Cr and Mn is revealed. The XMCD spectra (Figure 5) confirms the antiparallel spin alignment between Cr and Mn atom, as opposite signs of XMCD intensities arises at the $L_{2,3}$ edges. This antiparallel arrangement is a characteristic hallmark of ferrimagnetic ordering and the oppositely aligned atomic moments result in total magnetic moment compensation. The exchange coupling parameter plots shown in figure 6, further illustrate the microscopic origin of the antiparallel Cr and Mn spin alignment. The large negative $J_{ij}$ value of the inter-sublattice interaction between Cr and Mn indicates strong antiferromagnetic coupling. This drives the antiparallel alignment between Cr and Mn sublattices ultimately leading to the emergence of ferrimagnetism. Additionally, the $J_{ij}$ exchange plot also confirms the minimal long-range magnetic frustration that shows magnetic compensation is not a result of randomness but of well-defined competing exchange paths.

The antiparallel alignment between Cr (↓) and Mn (↑) sublattices leads to a spin-split band structure, where Cr-dominated minority-spin exhibit a gap, while Mn dominated majority spin show a robust metallic character due to partially filled states near the Fermi level. Additionally, strong $p$–$d$ hybridization between Mn-$d$ and Sb-$p$ orbitals near Fermi introduces in-gap states, further enhancing the metallicity of the spin-up channel. This asymmetric spin-resolved band filling, prevents the material from behaving as a true semiconductor. Furthermore, the antiferromagnetic super-exchange interaction, mediated via Sb-$p$ orbitals between Cr and Mn atoms, stabilizes the antiparallel configuration without enforcing a semiconducting gap. Although, magnetic compensation fulfils the Slater–Pauling criterion for net zero magnetic moment, the presence of localized sublattice moments leads to spin-polarized band renormalization, that enforces half-metallic ferromagnetism. Consequently, CrMnSb deviates from 18 electron non-magnetic semiconducting ground state, despite its ideal valence electron count.

**Authors Contribution**

**Himanshu Joshi**: Data curation (supporting); Formal analysis (lead); Investigation (lead); Methodology (equal); Validation (equal); Writing – original draft (lead). **Lalrin Kima, Shradhanjali Dewan, Aldrin Lalremtluanga, Homnath Luitel, K. C. Bhamu**: Data curation (supporting); Formal analysis (supporting); Investigation (supporting); Methodology (supporting); Validation (supporting). **D.P. Rai**: Data curation (lead); Formal analysis (supporting); Investigation (supporting); Methodology (equal); Validation (equal); Writing – original draft (supporting).

**Acknowledgement**

DPR acknowledges Science Engineering Research Board (SERB), New Delhi Govt. of India via File Number: SIR/2022/001150.

**Conflict of Interest**

The authors declare no conflict of interest.

**Data Availability**

The data that support the findings of this study are available from the corresponding author upon reasonable request

**Declaration of generative AI and AI-assisted technologies in the writing process**

During the preparation of this work the author(s) used ChatGPT (unpaid version) in order to improve language and readability. No AI tools were used to generate any text or images. After using this tool/service, the author(s) reviewed and edited the content as needed and take(s) full responsibility for the content of the publication.